\documentclass[10pt,twocolumn]{article}

\usepackage{braket}
\usepackage{bm}
\usepackage{amsmath}
\usepackage{graphicx}
\usepackage{pgf}
\usepackage{amsfonts}
\usepackage[hidelinks]{hyperref}
\usepackage{siunitx}
\usepackage{tabularx}
\usepackage{booktabs}
\usepackage{authblk}
\usepackage{floatpag}
\usepackage{caption}

\title{Managing uncertainty in data-derived densities to accelerate density functional theory}
\author[1]{Andrew T. Fowler}
\author[1,2]{Chris J. Pickard}
\author[1]{James A. Elliott}
\affil[1]{Department of Materials Science and Metallurgy, University of Cambridge, 27 Charles Babbage Road, Cambridge, CB3 0FS, United Kingdom}
\affil[2]{Advanced Institute for Materials Research, Tohuku University, 2-1 1 Katahira, Aoba, Sendai, 980-8577, Japan}
\date{}

\newcommand{\code}{\texttt}

\begin{document}

\maketitle
\begin{abstract}
Faithful representations of atomic environments and general models for regression can be harnessed to learn electron densities that are close to the ground state. One of the applications of data-derived electron densities is to orbital-free density functional theory. However, extrapolations of densities learned from a training set to dissimilar structures could result in inaccurate results, which would limit the applicability of the method. Here, we show that a non-Bayesian approach can produce estimates of uncertainty which can successfully distinguish accurate from inaccurate predictions of electron density. We apply our approach to density functional theory where we initialise calculations with data-derived densities only when we are confident about their quality. This results in a guaranteed acceleration to self-consistency for configurations that are similar to those seen during training and could be useful for sampling based methods, where previous ground state densities cannot be used to initialise subsequent calculations.
\end{abstract}

\section{Introduction}
Density functional theory (DFT) has seen widespread adoption in many areas of research spanning the natural sciences due to its high predictive capability at modest computational cost and transferability across different systems~\cite{Burke2012}. The staggering number of applications and papers that exploit DFT are a testament to its value in Materials Science~\cite{Jones2015}. 

The foundations of DFT are the Hohenberg-Kohn theorems~\cite{Hohenberg1964}. The first of these expresses the total energy of a many-electron system as a functional, $F[n]$, of the ground state electron density, $n(\mathbf{x})$, where $\mathbf{x}$ denotes a location in real space. The second theorem tells us the ground state density is found by minimising $F[n]$ with respect to $n(\mathbf{x})$. Although an exact form for $F[n]$ has not been established, the unknown components can be separated into a kinetic energy contribution, $T[n]$ and a term called the exchange correlation functional, $E_{\textnormal{xc}}[n]$~\cite{Cohen2012}. The magnitude of contributions from $E_{\textnormal{xc}}[n]$ to the total energy are known to be relatively small and so the exchange correlation term can be approximated to some extent by approaches like the local density approximation and the generalised gradient approximation~\cite{Kohn1965,Becke1986}. The kinetic energy term cannot however be so well approximated and a universally applicable functional is still unknown~\cite{Witt2018}. This forces many applications to an alternative paradigm, Kohn-Sham (KS) DFT~\cite{Kohn1965}. Here, $T[n]$ is replaced by an expectation over independant electron wave functions. In many cases, this vastly improves the accuracy of the kinetic energy contribution to $F[n]$ but it introduces a significant increase in the computational expense~\cite{Payne1992}.

With the recent renewed interest in machine learning, theoretical attempts to learn $T[n]$ have been supplemented with data-driven inferences~\cite{Snyder2012}. These are hampered by difficulties in approximating gradients $\partial T[n] / \partial n (\mathbf{x})$, an evaluation which is necessary in finding the ground state density~\cite{Li2016}. Recently, an approach to circumvent this issue was proposed, stimulating a new wave of interest in data-driven orbital free (OF) DFT~\cite{Brockherde2017,Sinitskiy2018,Grisafi2018}. The alternative route to evaluating data-derived OF functionals on the ground state density is to empirically infer the ground state density itself, removing the variational optimisation of $F[n]$ completely. Two possible issues with this approach ultimately stem from the availability of data. While $T[n]$ and $n(\mathbf{x})$ may be very accurate for structures similar to those seen during training, when extrapolating for unfamiliar structures, either $T[n]$ or $n(\mathbf{x})$ may give predictions that are far from the true values. 

The key contribution that we make in this work is to show that predictive uncertainty can be harnessed to prevent poor extrapolations of $n(\mathbf{x})$ for structures that are dissimilar to those seen during training. We illustrate how such a measure of confidence can be applied to accelerate KS DFT by initialising calculations with a data-driven contribution only when we are confident about its quality. We note that such an application is most suited to sampling methods such as nested sampling, where subsequent structures are not guaranteed to be similar~\cite{Brewer2011}. To the best of our knowledge, ab initio. nested sampling has yet to be realised due to the prohibitive computational requirements of standard KS DFT. This work may contribute, in some part, to realising such calculations. For other applications like molecular dynamics or geometry optimisation, a temporary history of ground state densities can be applied to subsequent configurations in the calculation. This results in successive calculations being initialised fairly close to their ground state, rendering any improvements made from a data-derived density to be much less significant.

\section{Quantifying uncertainty}
Evaluating an error or measure of confidence in a data-driven prediction like $n(\mathbf{x})$ is a well studied problem~\cite{Bishop2006,Ghahramani2015}. Applications of uncertainty quantification have recently begun appearing in Materials Science, with some even in DFT, such as the linear model exchange correlation functional of Aldegunde \emph{et al}. ~\cite{Aldegunde2016a,Aldegunde2016,Mo2018,Longbottom2018,Parks2018}. In this work, we show that useful applications of a predictive uncertainty in $n(\mathbf{x})$ can be realised for just one of many possible approaches. By illustrating a proof-of-concept application to accelerating KS DFT, we hope to encourage a greater awareness of the advantages of quantifying uncertainty and to stimulate interest in alternative methods and applications such as in OF DFT.

\subsection{Non-Bayesian regression}\label{sec:nonbayesregressions}
In the following we adopt the notation that $n$ and $\mathbf{x}$ refer to a known ideal model contribution to electron density and a corresponding representation for the environment of that density point, respectively. Specifically, we adopt the bispectrum representation for $\mathbf{x}=(\mathbf{x}^{\textnormal{local}},\mathbf{x}^{\textnormal{global}})$, which is a concatenation of local and global contributions~\cite{Bartok2010,Bartok2013}. We refer the reader to section \hyperref[sec:appendix_cnlm]{A} of the Appendix for further detail and also note that explicit dependence of $n$ upon $\mathbf{x}$ has been dropped in this section to improve clarity. In this work, we use a non-Bayesian approach to quantify uncertainty. Although a Bayesian method to parametric regression will give a more reliable measure of uncertainty, evaluating uncertainty from the predictive distribution for non-linear models is not a simple task and often sampling is involved which can incur significant computational overhead~\cite{Bishop2006}.

We propose a model in which observations of the true ground state density $n$ are prone to random error which is distributed normally about the model predictions $\mu(\mathbf{x},\mathbf{w})$:

\begin{equation}\label{eqn:likelihood_restrictedDFT}
p(n | \mathbf{x} , \mathbf{w}) = \mathcal{N}(n | \mu(\mathbf{x},\mathbf{w}),\sigma(\mathbf{x},\mathbf{w})^2) .
\end{equation}

We also introduce a dependency of the variance of this random error, $\sigma(\mathbf{x},\mathbf{w})^2$, on the environment $\mathbf{x}$, which is known as a \textit{heteroskedastic} model for noise~\cite{Pearson1896}. We use a fully connected feed-forward neural network with hidden network weights $\mathbf{w}$, to calculate $\mu(\mathbf{x},\mathbf{w})$ and $\sigma(\mathbf{x},\mathbf{w})^2$. Observations of $n$ are treated as independent and identically distributed random variables and to infer $\mathbf{w}$, we calculate the maximum likelihood estimate by maximising the product $\prod_{n,\mathbf{x}} p(n | \mathbf{x} , \mathbf{w})$ over all observations in the training set. 

To quantify error in the predictions of $n$ given a new $\mathbf{x}$, we adopt an ensemble of $N_{\textnormal{ens}}$ neural networks, each with network weights $\mathbf{w}_i$. Adopting a uniformly weighted Gaussian mixture, the likelihood of the ensemble is then: 

\begin{equation}\label{eqn:mixture_model_definition:restricted}
\begin{split}
p(n | \mathbf{x} , \mathbf{W}) &= \frac{1}{N_{\textnormal{ens}}} \sum_{i=1}^{N_{\textnormal{ens}}} \mathcal{N}\left(n | \mu(\mathbf{x},\mathbf{w}_i) , \sigma(\mathbf{x},\mathbf{w}_i)^2 \right) \\
&= \mathcal{N}\left(n | n^{\textnormal{\tiny{ML}}}(\mathbf{x},\mathbf{W}) , \sigma^{\textnormal{\tiny{ML}}}(\mathbf{x},\mathbf{W})^2 \right)
\end{split}
\end{equation}
where $\mathbf{W}=(\mathbf{w}_1,...,\mathbf{w}_{N_{\textnormal{ens}}})$ and $p(n | \mathbf{x} , \mathbf{W})$ is also a normal distribution~\cite{Lakshminarayanan2016}. Uncertainty in our prediction of $n$ is given by the variance of $p(n | \mathbf{x} , \mathbf{W})$, $\sigma^{\textnormal{\tiny{ML}}}(\mathbf{x},\mathbf{W})^2$ and can be evaluated as:

\begin{equation}\label{eqn:variance_eqn}
\begin{split}
\sigma^{\textnormal{\tiny{ML}}}(\mathbf{x})^2 &= \frac{1}{N_{\textnormal{ens}}}\sum_{i=1}^{N_{\textnormal{ens}}}\mu(\mathbf{x},\mathbf{w}_i)^2 - n^{\textnormal{\tiny{ML}}}(\mathbf{x})^2 \\
&+\frac{1}{N_{\textnormal{ens}}}\sum_{i=1}^{N_{\textnormal{ens}}} \sigma(\mathbf{x},\mathbf{w}_i)^2  \\
n^{\textnormal{\tiny{ML}}}(\mathbf{x}) &= \frac{1}{N_{\textnormal{ens}}}\sum_{i=1}^{N_{\textnormal{ens}}} \mu(\mathbf{x},\mathbf{w}_i) .
\end{split}
\end{equation}

\subsection{Doing no harm}
To apply our model for prediction uncertainty $\sigma^{\textnormal{\tiny{ML}}}(\mathbf{x})^2$ in \eqref{eqn:variance_eqn} to accelerate KS DFT, we need to evaluate a global measure of uncertainty for an entire structure. We call this measure $H[p_{\sigma^{\textnormal{\tiny{ML}}}}]$, where an unknown dependency on the empirical prior distribution $p_{\sigma^{\textnormal{\tiny{ML}}}}$ of $\sigma^{\textnormal{\tiny{ML}}}$ is shown explicitly. In this work, we adopt the very simple measure that:

\begin{equation}\label{eqn:statistic_H}
\begin{split}
H[p_{\sigma^{\textnormal{\tiny{ML}}}}] &= \mathbb{E}_{p_{\sigma^{\textnormal{\tiny{ML}}}}}[\ln(\sigma^{\textnormal{\tiny{ML}}})] \\
&= \frac{1}{N} \sum_{i=1}^N \ln(\sigma^{\textnormal{\tiny{ML}}}(\mathbf{x}_i))
\end{split}
\end{equation}
for $N$ densities in a crystal. We now abbreviate $H[p_{\sigma^{\textnormal{\tiny{ML}}}}]=H$ and introduce a tapering function $\Gamma(H)$, which is essentially a step function with a controllable transition point and length scale. For details of the specific form of $\Gamma$ used in this work, we refer the reader to section \hyperref[sec:appendix_calc_details]{C} of the Appendix. With $\Gamma(H)$, we can control the empirical contribution $n^{\textnormal{\tiny{ML}}}(\mathbf{x})$ to an initial density estimate:

\begin{equation}\label{eqn:density_model}
n(\mathbf{x}) = n_0(\mathbf{x}) + \Gamma(H) n^{\textnormal{\tiny{ML}}}(\mathbf{x}) .
\end{equation}
$n_0(\mathbf{x})$ represents any standard initialisation technique for the density in DFT but typically, this is a combination of the radial components of electron density for atoms assumed to be in vacuum. The ideal model contribution $n$ from section \ref{sec:nonbayesregressions} is the difference of the true ground state density and the standard initial contribution, $n(\mathbf{x})-n_0(\mathbf{x})$ from \eqref{eqn:density_model}.

We note that an alternative strategy could be to taper empirical contributions locally at each grid point, but we choose a global approach to discourage spurious non-smoothness in $n^{\textnormal{\tiny{ML}}}(\mathbf{x})\Gamma(\sigma^{\textnormal{\tiny{ML}}}(\mathbf{x}))$ that might occur if $|\sigma^{\textnormal{\tiny{ML}}}(\mathbf{x}+\delta\mathbf{x}) - \sigma^{\textnormal{\tiny{ML}}}(\mathbf{x})|>>0$, for a small perturbation in environment $\delta\mathbf{x}$. We also note that the effects of any random error in $\sigma^{\textnormal{\tiny{ML}}}(\mathbf{x})$ are significantly reduced by considering distribution averages. While we found that the simple choice of $H$ used in \eqref{eqn:statistic_H} worked very well at identifying uncertain predictions for the applications in this work, a more informative measure of the distribution $p_{\sigma^{\textnormal{\tiny{ML}}}}$ may improve this distinction further. Higher order moments of $p_{\sigma^{\textnormal{\tiny{ML}}}}$ such as the distribution variance for example could be utilised, in addition to knowledge about the distribution mean.

\section{Results}

%%%%%%%%%%%%%%%%%%% THIS FIGURE IS VERY SLOW TO COMPILE %%%%%%%%%%%%%%%%%
\begin{figure}
\centering
\input{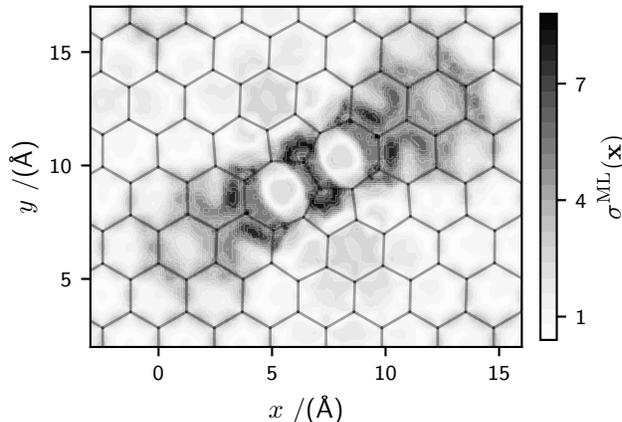}   %% UNCOMMENT for production (needs LuaLatex)
\caption{A model trained on pristine graphene identifies a large degree of prediction uncertainty, $\sigma^{\textnormal{ML}}(\mathbf{x})$ denoted by greyscale shading, in the area surrounding a 7-5 pair defect. We note that $\sigma^{\textnormal{ML}}(\mathbf{x})$ is given in units of $10^{-2} \si{\elementarycharge\angstrom}^{-3}$.}\label{fig:fig-data_3}
\end{figure}
%%%%%%%%%%%%%%%%%%% THIS FIGURE IS VERY SLOW TO COMPILE %%%%%%%%%%%%%%%%%
In this section we illustrate how the non-Bayesian approach to uncertainty quantification adopted in this work can qualitatively distinguish accurate from inaccurate values of the data-derived contribution $n^{\textnormal{\tiny{ML}}}(\mathbf{r})$. We also show how the number of self-consistent field iterations needed to reach self-consistency in a KS DFT calculation can be reduced as the initial density tends to the exact ground state density. 

\vspace{2mm}
For environments dissimilar to those seen during training, we expect a larger predictive uncertainty.
\paragraph{7-5 defect in graphene}
 Figure~\ref{fig:fig-data_3} shows $\sigma^{\textnormal{\tiny{ML}}}(\mathbf{x})$ for a single layer of graphene with a 7-5 pair (Dienes) topological defect~\cite{Monthioux2014}. Only densities from a single pristine layer of graphene were used during training. In the area immediately surrounding the defect, predictive uncertainties increase (denoted by dark shading), identifying this region as an environment dissimilar to the defect-free layer.

\begin{figure}
\centering
\input{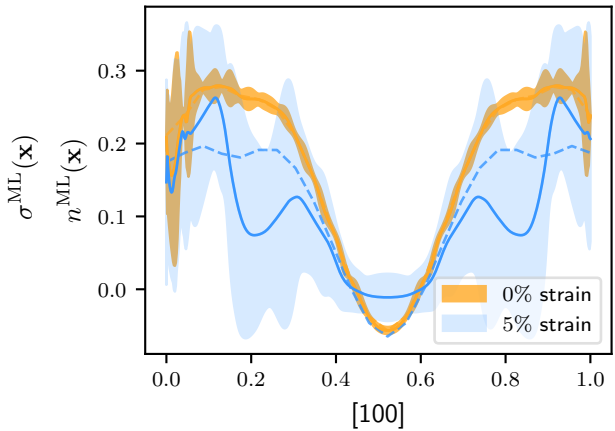}
\caption{A model trained only on primitive graphite with 0\% in-plane lattice strain identifies a region of high degree of uncertainty when making predictions along the [100] contour of a primitive graphite crystal with 5\% in-plane strain. The shaded regions show the interval $n^{\textnormal{\tiny{ML}}}(\mathbf{r})\pm3\sigma^{\textnormal{\tiny{ML}}}(\mathbf{x})$ and the dashed lines show the true ground state density. We note that charge densities are given in units of \si{\elementarycharge\angstrom}$^{-3}$.}\label{fig:fig-data_1}
\end{figure}

\paragraph{In-plane strain in graphite}
In Figure~\ref{fig:fig-data_1}, we compare the prediction uncertainty of graphite with 0\% and 5\% in-plane strain. Specifically, we show the [100] lattice vector contour and find that predictions are significantly more certain for the 0\% contour which was seen during training, than the 5\% contour that was not. Further details of the bispectrum and KS DFT calculations for Figures~\ref{fig:fig-data_3} and~\ref{fig:fig-data_1} can be found in the Appendix, section \hyperref[sec:appendix_calc_details]{C}.

\subsection{Accurate initial densities}
To motivate our application of uncertainty quantification to KS DFT, we examine the convergence of single point KS DFT calculations to self-consistency, as we perturb initial densities away from the exact ground state via perturbations to the ideal model contribution, $n(\mathbf{x})-n_0(\mathbf{x})$ in \eqref{eqn:density_model}.

We study a non-metallic crystal, graphite, and calculate the ground state density for several hundred primitive cell configurations sampled from a NPT molecular dynamics trajectory. The components of the discrete Fourier transform of the ideal model contribution are perturbed by additive Gaussian noise. By taking the inverse transform, we have a continuously deformed version of the true density. We measure deformation by the root-mean-squared error ($\textnormal{RMSE}$) of the perturbed and true ground state density.
                    
\begin{figure}
\centering
\input{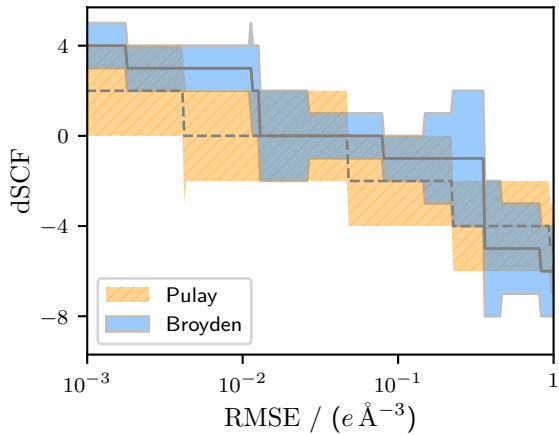}
\caption{As the initial KS DFT densities are deformed away from the true ground state, the number of iterations necessary to reach self-consistency increases and the improvement from standard DFT initialisation, $\textnormal{dSCF}$, decreases. Pulay and Broyden density mixing schemes result in a very similar convergence for all calculations.}\label{fig:fig-data_2}
\end{figure}

In Figure~\ref{fig:fig-data_2}, the self-consistent calculation is initialised with charge densities which are increasingly deformed versions of the ground state. We see that as the magnitude in deformation from the ground state, the $\textnormal{RMSE}$, increases, so too does the number of iterations needed to reach self-consistency.  The quantity displayed on the abscissa, $\textnormal{dSCF}$, is the improvement, in the number of iterations, relative to a calculation with the standard initial density. As deformations increase, the improvement decreases. The hashed and shaded areas in Figure~\ref{fig:fig-data_2} represent confidence intervals of 67\%, showing that the relation between $\textnormal{RMSE}$ and convergence to self-consistency is stochastic to some degree. 

To ensure that any empirical method for initialising KS DFT densities does not negatively affect convergence to self-consistency in regions where the empirical densities extrapolate poorly, uncertainty quantification is clearly needed. For further details of the DFT calculations in Figure~\ref{fig:fig-data_2}, see section \hyperref[sec:appendix_calc_details]{C} of the Appendix.

\section{Applying global uncertainty}
As we saw by the calculations in Figure~\ref{fig:fig-data_2}, a measure of confidence in density is necessary if we are to use empirical densities in DFT in a ``safe'' manner. Not wanting to leave things worse than how we found them, we hope to ensure that every calculation initialised by a data-driven density does no worse than its ordinary counterpart.

\begin{figure}
\input{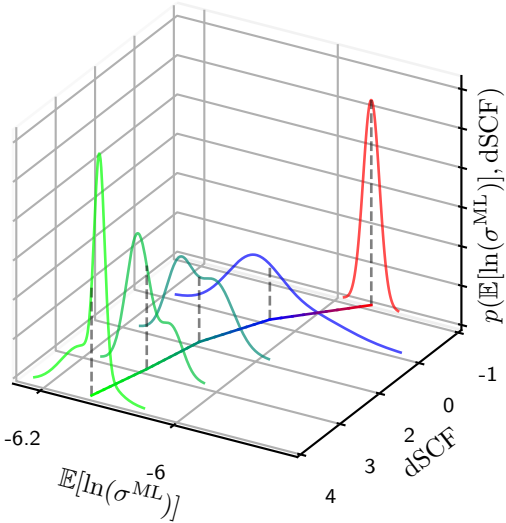}
\caption{The empirical joint distribution $p(H , \textnormal{dSCF})$ evaluated for a set of primitive graphite configurations shows separable categorisations of performance into good ($\textnormal{dSCF}>0$) and poor ($\textnormal{dSCF}<0$) predictions.}\label{fig:fig-data_4}
\end{figure}

To illustrate that a global measure of confidence in predictions, $H$ from \eqref{eqn:statistic_H}, can be applied to accelerate KS DFT, we first consider using empirical densities without using knowledge of their uncertainty. After training on 5 primitive cell graphite configurations from a NVT molecular dynamics simulation at $T=\SI{300}{\kelvin}$, we predict densities for all 300 crystals in our data set. Details of the empirical model and KS DFT calculations can be found in the Appendix, section \hyperref[sec:appendix_calc_details]{C}. Without applying any information about uncertainty, we blindly initialise Broyden density mixing (DM) DFT calculations and record the reduction in the number of iterations to self-consistency, $\textnormal{dSCF}$, relative to a calculation with a standard initial density. Next, we calculate the global confidence measure $H=\mathbb{E}[\ln(\sigma^{\textnormal{\tiny{ML}}})]$ for each crystal and categorise crystals into discrete sets according to their $\textnormal{dSCF}$ score. We show the corresponding empirical joint distribution $p(\mathbb{E}[\ln(\sigma^{\textnormal{\tiny{ML}}})] , \textnormal{dSCF})$ in Figure~\ref{fig:fig-data_4}.

We can expand the empirical joint distribution 

\begin{equation}
p(H, \textnormal{dSCF}) = p(\textnormal{dSCF} | H) p(H)
\end{equation}

in terms of the unknown conditional distribution $p(\textnormal{dSCF} | H)$ from which we want to decide if a given prediction, $H$ is good enough to initialise a KS DFT calculation. Taking the prior $p(H)$ as constant, $p(H, \textnormal{dSCF}) \propto p(\textnormal{dSCF} | H)$ and we look for a gap in $H$ between $p(H, \textnormal{dSCF} \ge 0)$ and $p(H, \textnormal{dSCF} < 0)$. It is here that a transition point can be set in the tapering function $\Gamma(H)$, to reduce uncertain predictions to zero. This can be visualised by comparing the peak at $\textnormal{dSCF}=-1$ with that at $\textnormal{dSCF=0}$. Although there is some overlap between these two peaks, there is almost zero overlap between $\textnormal{dSCF=-1}$ and all other peaks. This means that $H$ can be used to identify the quality of density predictions before any DFT calculation is made. We note that the joint distribution shown in Figure~\ref{fig:fig-data_4} is a smoothed approximation of the true empirical distribution but that important properties such the width of each conditional distribution $p(\textnormal{dSCF} | H)$, are preserved.

In fact we see that for the small study here, the expectation of $H$ conditioned on $\textnormal{dSCF}$,

\begin{equation}
\mathbb{E}_{p(H | \textnormal{dSCF})}[H] = \int \textnormal{d}H  p(H | \textnormal{dSCF}) H
\end{equation}

which is the dashed line in Figure~\ref{fig:fig-data_4}, follows a monotonic relation with $\textnormal{dSCF}$. This shows that predicted uncertainty really does correspond in a monotonic way to actual error. Using the distribution of predicted uncertainties over a crystal, we can identify model predictions which are poor and will harm converge to self-consistency. By effectively turning off poor predictions using $\Gamma(H)$, empirical corrections to the initial KS density can be applied only for crystals which are similar to those seen during training.

\subsection{Accelerating self-consistency}
\begin{figure}
\input{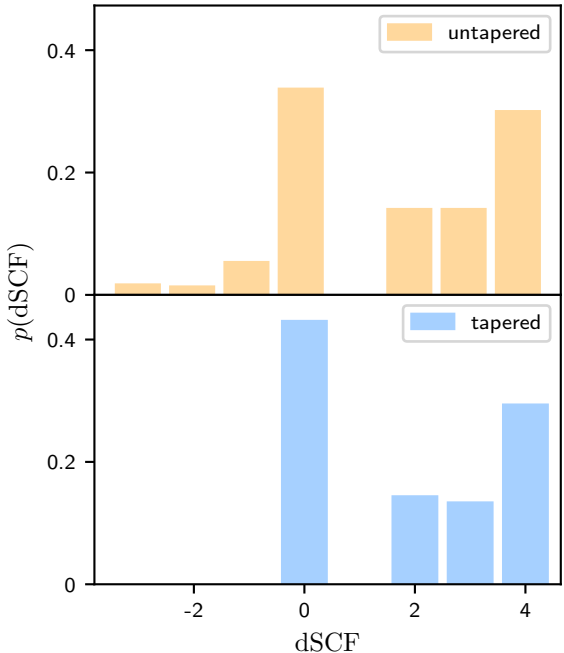}
\caption{When data-derived densities are used in KS DFT without using uncertainty prediction (untapered contributions), there is a non-zero chance $p(\textnormal{dSCF})$ that inaccurate predictions can harm convergence to self-consistency ($\textnormal{dSCF}<0$). When prediction uncertainties are applied to identify (taper) unfamiliar crystals, only a neutral or positive speed up is seen.}
\label{fig:fig-data_5}
\end{figure}

Now that we have established a mechanism to detect global uncertainty in density, we can apply this to single point KS DFT calculations to accelerate convergence to self-consistency. In Figure~\ref{fig:fig-data_5} we compare the empirical distributions $p(\textnormal{dSCF})$ for Broyden DM DFT calculations performed using data-derived densities with and without tapering. The upper half of Figure~\ref{fig:fig-data_5} shows a number of extrapolations where poor predictions of density have a negative effect upon convergence ($\textnormal{dSCF}<0$). In the lower half, uncertain predictions have been identified and reduced to zero, increasing the peak at $\textnormal{dSCF}=0$. 

Crucially, the computation time required to evaluate our data-derived density estimate is just less than the time taken to evaluate a single SCF iteration. For further details of the calculations in Figure~\ref{fig:fig-data_5} involving KS DFT parameters, see section \hyperref[sec:appendix_calc_details]{C} of the Appendix. Despite our model being trained only on 5 configurations, a large proportion of crystals exhibit a speed up in converging to self-consistency. Such an effect could also arise by poorly choosing a test set of crystals, whereby all atom positions remain in almost identical positions. A trivial approach of applying the ground state density from a random crystal, or an average of ground state densities over all crystals, would therefore achieve similar, or better results. However, in fact this is not the case. When such simulations were run, we found that almost all ($\sim95\%$) of predictions obtained $\textnormal{dSCF}=0$. Our test set is in fact a rather dissimilar collection of configurations, most of which involve significant shifts in registry across the basal plane, as configurations jump from one AB-stacked state to another. We attribute the ability of our model to infer useful predictions from such an incredibly small number of configurations to the fact that each crystal in the training set contributes $\mathcal{O}(10^4)$ grid points. Simply put, more data leads to a better inference, even when a large number of data points come from the same crystal.

\subsection{Wider applicability}\label{sec:wider_applicability}
To this point, all calculations in this work have been made to illustrate that data-derived densities can be applied to a single system to improve the standard analytical initial densities that are used in KS DFT. We consider the wider implications of this work beyond graphite by comparing the dissimilarity of ground state and standard initial densities for a collection of 29 metals and 37 non-metals under both low and high pressure. We find that all of the metals we consider have initial densities that are much closer to the ground state density than with graphite while the converse is true for approximately half of the non-metals studied here. We use the RMSE of all density grid points within a crystal as a measure of dissimilarity between these two densities. To classify metals and non-metals we use the density of states at the Fermi level. We use a value of $\SI{0.2}{\elementarycharge(\electronvolt)^{-1}}$ which is just above the density of states for the metalloid As to classify the two classes.

\begin{figure}
\input{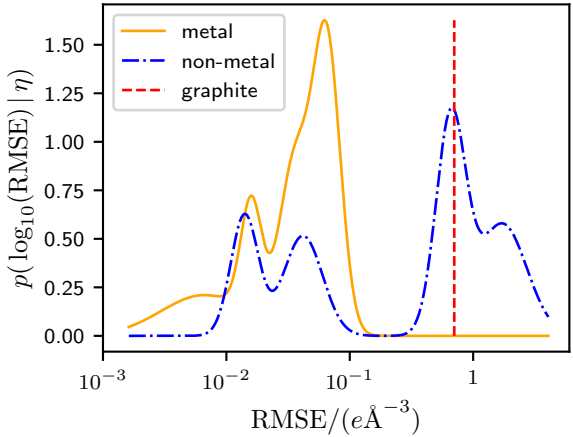}
\caption{The $\textnormal{RMSE}$ between the standard initial and ground state densities is much smaller for the metals studied here then with graphite and approximately half of the non-metals. Four-component Gaussian mixture models approximate the conditional distributions $p(\log_{10}(\textnormal{RMSE})|\eta)$ of the $\textnormal{RMSE}$ given the characterisation $\eta$ of the system. The vertical dashed line shows the RMSE of graphite.}\label{fig:fig-data_6}
\end{figure}

We show in Figure~\ref{fig:fig-data_6} a smoothed approximation of the conditional distribution $p(\hspace{0.5mm}\log_{10}(\textnormal{RMSE})\hspace{0.5mm}|\hspace{0.5mm}\eta)$ for metal or non-metals $\eta$ along with a dashed vertical line showing the $\textnormal{RMSE}$ between the standard initial and ground state densities for graphite. The logarithm of the $\textnormal{RMSE}$ illustrates that the $\textnormal{RMSE}$ differs by almost two orders of magnitude between graphite and some of the metals. We note that the approximate representation of $p(\hspace{0.5mm}\log_{10}(\textnormal{RMSE})\hspace{0.5mm}|\hspace{0.5mm}\eta)$ is a four-component Gaussian mixture model of the true data~\cite{Bishop2006}. Based upon the systems studied here, we summarise that data-derived densities may in general be more suitably applied to non-metals than metals. Details of these systems as well as and the KS DFT calculations that were used to calculate the $\textnormal{RMSE}$ in Figure~\ref{fig:fig-data_6} can be found in section \hyperref[sec:appendix_calc_details]{C} of the Appendix.

\section{Discussion}
We have shown that uncertainty quantification can be applied to accelerate KS DFT for sampling methods like nested sampling, attaining a maximum speed up of 57\%\footnote{See section \hyperref[sec:appendix:acceleration]{D} of the Appendix for the definition of speed up that we adopt here.\label{footnotelabel}} for the systems studied in this work. We view the approach taken in this work as more a proof of concept than a final solution, confident that exciting developments and insights are accessible to future work. To this end, we note that the approach taken in this work is just one of many possible methods. We use this section to discuss what we believe to be the most prominent disadvantages of this approach and outline a few ideas that could address these short comings.

While our parametric approach leads to a computation time for data-derived densities that is smaller than a single self-consistent field cycle, the time needed to train densities from a single crystal is orders of magnitude larger than a full DFT calculation. Although sampling methods require thousands of crystal configurations, the time to train or refine an data-derived density should ideally be as close to a single KS DFT calculation as possible. Some heuristic techniques, such as maximising the sample variance of observations in a smaller training subset, may give some reduction in this, but a more promising avenue could be to use approximate Bayesian inference, such as deterministic variational inference~\cite{Wu2018}. A Bayesian approach, even when the posterior distribution is approximate, could also lead to more reliable uncertainty estimates. The non-Bayesian approach adopted in this work does not guarantee that ``false positives'' cannot occur when determining if confidence should be placed in a data-derived density or not. In addition, Bayesian online learning could allow for an incremental approach to learning densities such that refinements are made during sampling, only to crystals which are dissimilar to all of those that were previously seen during training~\cite{Opper1999}.

The approach that we use in this work to make decisions about confidence in the density, does not take account of the type of crystal. For applications like nested sampling where several different phases are sampled from, it may become essential for our decision process to include knowledge about the global environment, such as from a global bispectrum representation of the crystal. An unsupervised method such as a Gaussian mixture model may be necessary to associate crystals with nearby clusters and to apply decisions using a predetermined set of distinct confidence thresholds for each cluster. 

An aspect that we haven’t considered in this work is the question of which method of minimising the KS Hamiltonian, given an initial density, gives the lowest computation time. Although this is a well studied problem, perhaps new insights are possible when an estimate of confidence is available in the initial density~\cite{Woods2018}. %One might expect for example, that very close to the ground state, ensemble DFT might be preferable over DM schemes~\cite{Marzari1997}.

%\subsubsection{Unrestricted spin}
We note that our discussion of KS DFT and the application of data-derived densities to accelerate convergence to self-consistency in this work has so far ignored spin. For many systems and processes such as radicals, transition metal complexes or homolytic bond breaking, the spatial wave functions of opposing spin states are not equal ${(\psi^{\alpha}(\mathbf{r})\neq\psi^{\beta}(\mathbf{r}))}$~\cite{Grafenstein2002,Power2012,Yamanaka1994}. Spin-unrestricted KS DFT is a generalisation of the spin-restricted form, where $\psi^{\alpha}(\mathbf{r})\neq\psi^{\beta}(\mathbf{r})$ is possible and the variational minimisation of total energy $E[n,Q]$ is performed with respect to both the total electron density $n(\mathbf{r})$ and the spin density ${Q(\mathbf{r})=\sum_i|\psi^{\alpha}(\mathbf{r})|^2-\sum_i|\psi^{\beta}(\mathbf{r})|^2}$~\cite{Jacob2012}. Initial densities for unrestricted spin therefore require $Q(\mathbf{r})$ in addition to $n(\mathbf{r})$. A generalisation of the data-derived densities used in this work to unrestricted spin DFT could be realised by adopting ${p(\mathbf{t}|\mathbf{x},\mathbf{w}) = \mathcal{N}(\mathbf{t}|\bm{\mu},\bm{\Lambda}^{-1})}$ for ${\mathbf{t}=(n(\mathbf{r}),Q(\mathbf{r}))}$. A parametric model would then represent ${\mathbf{x}\rightarrow\left(\bm{\mu},\bm{\Lambda}^{-1}\right)}$ rather then ${\mathbf{x}\rightarrow(\mu,\sigma^2)}$ as in~\eqref{eqn:likelihood_restrictedDFT}. The generalisation of \eqref{eqn:mixture_model_definition:restricted} leads to ${p(\mathbf{t}|\mathbf{x},\mathbf{W})=\mathcal{N}(\bm{t}|\bm{\mu}^{\textnormal{\tiny{ML}}},(\bm{\Lambda}^{\textnormal{\tiny{ML}}})^{-1}}$ where ${\bm{\mu}^{\textnormal{\tiny{ML}}}=(n^{\textnormal{\tiny{ML}}},Q^{\textnormal{\tiny{ML}}})}$ and the covariance matrix $(\bm{\Lambda}^{\textnormal{\tiny{ML}}})^{-1}$ represents uncertainty in the initial data-derived total $(n^{\textnormal{\tiny{ML}}})$ and spin $(Q^{\textnormal{\tiny{ML}}})$ densities. The simplest way to apply $\bm{\Lambda}^{\textnormal{\tiny{ML}}}$ to identify uncertain predictions might be to sum the diagonal components of $(\bm{\Lambda}^{\textnormal{\tiny{ML}}})^{-1}$ to define a scalar measure analogous to $\sigma^{\textnormal{\tiny{ML}}}$ in~\eqref{eqn:statistic_H}. We also note that $E[n,Q]$ is well known to exhibit a number of stationary points and in the absence of any knowledge about the ground state of $Q$, some form of approximate global optimisation must be utilised. If the data-derived densities are sufficiently accurate then global optimisation for spin-unrestricted DFT could be abandoned altogether, providing significant reductions to the computation required.

\section{Conclusions}
We have shown that a non-Bayesian treatment of predictive uncertainty can be applied to electron density regression to identify crystals that are dissimilar to those seen during training. We have applied this approach to KS DFT where we have been able to identify and prevent unfamiliar crystals from negatively effecting convergence to self-consistency. For the systems studied in this work, where confident predictions were made we saw a maximum speed up in convergence to self-consistency of $57 \%$\textsuperscript{\ref{footnotelabel}} and cautiously note that further improvements could be made with a more in depth study of the approach to minimise the KS Hamiltonian. Crucially, our predictions can be evaluated in less time than a single self-consistent field iteration for a primitive crystal, meaning that our application to KS DFT could be useful for methods like nested sampling. 

We view this work as a proof of concept. Quantifying uncertainty in predicted densities is shown to be a fruitful endeavour and we hope our work will encourage further applications and alternative approaches, for example in orbital free DFT. More generally, this work motivates more sophisticated treatments of interpolation, or caching, which are currently treated deterministically to accelerate high performance plane wave DFT codes \cite{Clark2005,Hafner2008}. We anticipate that a paradigm shift towards ``probabilistic caching'', or regression, will lead to the efficient use of previously computed data.

\section*{Acknowledgements}
The authors would like to thank Nick Woods for reading an earlier version of this manuscript and for sharing a number of insightful observations regarding spin-polarised DFT that motivate our discussion of extending data-derived densities to spin-unrestricted DFT. We also thank Georg Schusteritsch for many helpful discussions regarding KS DFT and the self-consistent field procedure as well as our referees for providing constructive comments that has helped to improve this manuscript. Additionally, we thank the UKCP consortium, grant number EP/P022596/1 and the Royal Society through a Royal Society Wolfson Research Merit award, on behalf of C.J.P. . We also thank the EPSRC on behalf of A.T.F under the EPSRC Centre for Doctoral Training in Computational Methods for Materials Science, grant number EP/L015552/1. The plane wave code \code{CASTEP} was used used for all DFT calculations in this work \cite{Clark2005}. We provide our code for bispectrum and neural network calculations in the form of a Python module at \href{https://github.com/andrew31416/densityregression}{https://github.com/andrew31416/densityregression}.

\bibliography{main.bbl} 

\onecolumn
\twocolumn

\section*{Appendix}
\subsection*{A}\label{sec:appendix_cnlm}
In the bispectrum approximation, elements of local and global contributions to the representation of environment, $\mathbf{x}(\mathbf{r})$, are determined by the projections $c_{nlm}$ of local and global environment into radial ($n$) and spherical harmonic ($lm$) bases.

\begin{equation}
\begin{split}
c^{\textnormal{local}}_{nlm}(\mathbf{r}) &= \sum_{i \in \Omega_{\mathbf{r}}} g_n(\textnormal{d}r_i) Y_{lm}(\textnormal{d}\theta_i,\textnormal{d}\phi_i) \\
c_{nlm}^{\textnormal{global}} &= \frac{1}{N}\sum_{i=1}^N \sum_{j \in \Omega_{\mathbf{r}_i}} g_n(\textnormal{d}r_{ij})Y_{lm}(\textnormal{d}\theta_{ij},\textnormal{d}\phi_{ij}) \hspace{1mm}.
\end{split}
\end{equation}
$\Omega_{\mathbf{r}}$ is a spherical volume of finite radius surrounding a point in real space $\mathbf{r}$. Indices $i$ and $j$ denote pairs of the $N$ atoms contained in the primitive cell of a crystal. $\textnormal{d}r_i=|\mathbf{r}_i-\mathbf{r}|$, $\textnormal{d}r_{ij}=|\mathbf{r}_j-\mathbf{r}_i|$ and $\textnormal{d}\theta$ and $\textnormal{d}\phi$ are the polar and azimuth projections respectively of $\mathbf{r}_i-\mathbf{r}$ and $\mathbf{r}_j-\mathbf{r}_i$.

\begin{table}[b]
\begin{minipage}{1\textwidth}
\caption{Calculation parameters. }
\centering
\begin{tabular}{l || c c c | c c | c c}
\toprule\toprule
& \multicolumn{3}{c|}{bispectrum} & \multicolumn{2}{c|}{Neural network} & \multicolumn{2}{c}{KS DFT}\\
Calculation & $r_{\textnormal{cut}}$ (\si{\angstrom}) & $n_{\textnormal{max}}$ & $l_{\textnormal{max}}$ & $N_{\textnormal{ens}}$ & $\textnormal{nodes}$ & $E_{\textnormal{cut}}$ (\si{\electronvolt}) & $k$-point grid \\
\midrule
Figure~\ref{fig:fig-data_3} & 4 & 3 & 3 & 5 & $2\times100$ & 400 & [36 36  1] \\
Figure~\ref{fig:fig-data_1} & 6 & 10$^*$ & 8$^*$ & 10 & $2\times50$ & 400 & [36 36 6] \\
Figure~\ref{fig:fig-data_2} & - & - & - & - & - & 800 & [36 36 6] \\
Figure~\ref{fig:fig-data_4} & 6 & 6 & 6 & 10 & $2\times200$ & 300 & [18 18 4] \\
Figure~\ref{fig:fig-data_5} & 4 & 4 & 4 & 5 & $2\times150$ & 300 & [18 18 4] \\
Figure~\ref{fig:fig-data_6} & - & - & - & - & -            & 800 & (0.1,0.1,0.1)$\si{\angstrom^{-1}}$ \\
\bottomrule
\end{tabular}\label{tbl:appendix}
\end{minipage}
\end{table}

\subsection*{B}\label{sec:appendix_tapering}

There are many equally adequate tapering functions which could be chosen. We used:

\begin{equation} \label{eqn:appendix_tapering}
\begin{split}
\Gamma(\sigma(\mathbf{r}) , \sigma^*) = 
\begin{cases}
\frac{\tilde{\sigma}^4}{1+\tilde{\sigma}^4} &, \tilde{\sigma}>0 \\
0 &,\tilde{\sigma}<0
\end{cases} \\
\tilde{\sigma} = \frac{\sigma^* - \sigma(\mathbf{r})}{\sigma_{\textnormal{scale}}}
\end{split}
\end{equation}

simply because it has property that every $n^{\textnormal{th}}$ derivative $\partial^{(n)} \Gamma / \partial \sigma(\mathbf{r})^{(n)}$ remains continuous.

\subsection*{C}\label{sec:appendix_calc_details}

For all of the KS DFT calculations in Table~\ref{tbl:appendix}, a Fermi surface smearing width of 250 \si{\kelvin}, an energy tolerance of $10^{-6}$ \si{\electronvolt}/atom, the PBE exchange correlation functional and Broyden density mixing were used, except for the calculations in Figure~\ref{fig:fig-data_2} which used both Broyden and Pulay density mixing and the calculations in Figure~\ref{fig:fig-data_6} which used a smearing width of \SI{300}{\kelvin}. All graphite and graphene configurations except for the NPT calculations of figure \ref{fig:fig-data_2} had an in-plane C-C spacing of 1.42 \si{\angstrom} and an inter-layer spacing of 3.34 \si{\angstrom}. In addition, a vacuum corresponding to a unit cell of 20 \si{\angstrom} in the plane-normal axis was adopted for the graphene layer in Figure~\ref{fig:fig-data_3} and in Figure~\ref{fig:fig-data_5} a tapering function of the form in \eqref{eqn:appendix_tapering} was used with the threshold and scaling factor $(\sigma^*,\sigma_{\textnormal{scale}})=(-6.83,10^{-3})$. $*$ denotes use of the power spectrum rather than bispectrum representation for the calculations in Figure~\ref{fig:fig-data_1}. The notation adopted in \ref{tbl:appendix} regarding the number of nodes used in each neural network, is that $x\times y$ denotes a neural network of $x$ node layers, each containing $y$ nodes. Table~\hyperref[tbl:appendix:codicsd]{2} lists the database, unique identifying number and characterisation of each system used to generate the calculations in Figure~\ref{fig:fig-data_6}. We supply input files for all data sets within this work at \newline\href{https://github.com/andrew31416/densityregression/tree/master/data\_sets}{https://github.com/andrew31416/densityregression/}\newline\href{https://github.com/andrew31416/densityregression/tree/master/data\_sets}{tree/master/data\_sets}.

\begin{table}
\vspace{-3cm}
\begin{minipage}{1\textwidth}
\centering
\captionsetup{width=14cm}
\caption{The collection of metals and non-metals from Figure~\ref{fig:fig-data_6} were taken from the Crystallography Open Database (COD)~\cite{Grazulis2012} and the Inorganic Crystal Structure Database (ICSD)~\cite{Hellenbrandt2004}. A unique identifying number is given for each crystal along with its characterisation as discussed in section~\ref{sec:wider_applicability}}\label{tbl:appendix:codicsd}
\thisfloatpagestyle{empty}%
%auto-ignore
\begin{center}
\begin{tabular}{c|c|c||c|c|c}
\toprule\toprule
Database & Identifier & Characterisation &  Database & Identifier &Characterisation \\
\midrule
COD & 9008572 & non-metal&COD & 9008531 & metal \\
COD & 9008594 & non-metal&COD & 9008468 & metal \\
COD & 9008595 & non-metal&COD & 9008544 & metal \\
COD & 9008569 & non-metal&COD & 9008482 & metal \\
COD & 9008577 & non-metal&COD & 9008478 & metal \\
COD & 9008561 & non-metal&COD & 9008485 & metal \\
COD & 1011098 & non-metal&COD & 9008463 & metal \\
COD & 9008568 & non-metal&COD & 9008458 & metal \\
ICSD & 193853 & non-metal&COD & 9008552 & metal \\
ICSD & 26158 & non-metal&COD & 9008501 & metal \\
ICSD & 9863 & non-metal&COD & 9008490 & metal \\
ICSD & 18154 & non-metal&COD & 9008522 & metal \\
ICSD & 16516 & non-metal&COD & 9008470 & metal \\
ICSD & 15598 & non-metal&COD & 9008549 & metal \\
ICSD & 41440 & non-metal&COD & 9008513 & metal \\
ICSD & 2130 & non-metal&COD & 9008536 & metal \\
ICSD & 411857 & non-metal&COD & 9008514 & metal \\
ICSD & 27249 & non-metal&COD & 9008546 & metal \\
ICSD & 20904 & non-metal&COD & 9008584 & metal \\
ICSD & 22156 & non-metal&COD & 9008570 & metal \\
ICSD & 84461 & non-metal&COD & 9008543 & metal \\
ICSD & 15390 & non-metal&COD & 9008525 & metal \\
ICSD & 39566 & non-metal&COD & 9008512 & metal \\
ICSD & 27798 & non-metal&COD & 9008557 & metal \\
ICSD & 40914 & non-metal&COD & 9008477 & metal \\
ICSD & 16428 & non-metal&COD & 9008558 & metal \\
ICSD & 165592 & non-metal&ICSD & 15535 & metal \\
ICSD & 22157 & non-metal&ICSD & 63670 & metal \\
ICSD & 19079 & non-metal&ICSD & 653014 & metal \\
ICSD & 29128 & non-metal& & & \\
ICSD & 107946 & non-metal& & & \\
ICSD & 60559 & non-metal& & & \\
ICSD & 16262 & non-metal& & & \\
ICSD & 187642 & non-metal& & & \\
ICSD & 22158 & non-metal& & & \\
ICSD & 18012 & non-metal& & & \\
ICSD & 77378 & non-metal& & & \\
\bottomrule
\end{tabular}\label{tbl:appendix:datasetref}
\end{center}

\end{minipage}
\end{table}

\onecolumn
\twocolumn

\subsection*{D}\label{sec:appendix:acceleration}
We adopt the convention that the speed up 

\begin{equation}
\tau = \left(\frac{N_{\textnormal{old}} - N_{\textnormal{new}} }{N_{\textnormal{new}}}\right)\times 100\%    
\end{equation}
for data-derived initial densities that require $N_{\textnormal{new}}$ self-consistent field iterations to reach self-consistency. $N_{\textnormal{old}}$ is the number of iterations required for a standard calculation that uses a non data-derived initial density. $\tau$ is defined such that a data-derived density that halves then required number of self-consistent field iterations corresponds to a 100\% speed up.

\end{document}